\documentclass[conference]{IEEEtran}
\usepackage[utf8]{inputenc}
\usepackage[T1]{fontenc}
\usepackage{CJKutf8}
\usepackage{graphicx}
\usepackage{cite}
\usepackage{amsmath,amssymb,amsfonts}
\usepackage{algorithmic}
\usepackage{textcomp}
\usepackage{xcolor}
\usepackage{listings}
\usepackage{url}
\usepackage{booktabs,siunitx}
\usepackage{array,threeparttable,pifont}
\def\BibTeX{{\rm B\kern-.05em{\sc i\kern-.025em b}\kern-.08em
    T\kern-.1667em\lower.7ex\hbox{E}\kern-.125emX}}

\begin{document}

\title{Enterprise Data Science Platform: A Unified Architecture for Federated Data Access}

\author{\IEEEauthorblockN{Ryoto Miyamoto}
\IEEEauthorblockA{
\textit{Waseda University}\\
Tokyo, Japan \\
r-miyamoto@toki.waseda.jp}
\and
\IEEEauthorblockN{Akira Kasuga}
\IEEEauthorblockA{
\textit{CyberAgent, Inc.}\\
Tokyo, Japan \\
kasuga\_akira@cyberagent.co.jp}
}

\maketitle

\begin{abstract}
Organizations struggle to share data across departments that have adopted different data analytics platforms. In the Big Data era, organizations need to establish data as a shared asset accessible to various applications. When n datasets need to work with m analytical environments, this can create up to n×m replicas, leading to data inconsistencies and increased management costs. Traditional data warehouses require copying data into vendor-specific environments, while direct data access across different analytics platforms remains technically challenging. 
This study proposes the Enterprise Data Science Platform (EDSP), which builds on data lakehouse architecture and follows a Write-Once, Read-Anywhere principle. EDSP enables federated data access for multi-query engine environments, targeting data science workloads with periodic data updates and query response times ranging from seconds to minutes. By providing centralized data management with federated access from multiple query engines to the same data sources, EDSP eliminates data duplication and vendor lock-in inherent in traditional data warehouses. The platform employs a four-layer architecture: Data Preparation, Data Store, Access Interface, and Query Engines. This design enforces separation of concerns and reduces the need for data migration when integrating additional analytical environments.
Experimental results demonstrate that major cloud data warehouses and programming environments can directly query EDSP-managed datasets. We implemented and deployed EDSP in an enterprise production environment, confirming interoperability across multiple query engines. For data sharing across different analytical environments, EDSP achieves a 33–44\% reduction in operational steps compared with conventional approaches requiring data migration (e.g., Hugging Face Dataset Hub, BigQuery/Snowflake native tables). Although query latency may increase by up to a factor of 2.6 compared with native tables, end-to-end completion times remain on the order of seconds, maintaining practical performance for analytical use cases. Based on our production experience, EDSP provides practical design guidelines for addressing the data-silo problem in multi-query engine environments.
\end{abstract}


\section{Introduction}

When organizations need to access $n$ datasets from $m$ different analytics platforms, they often resort to creating up to $n \times m$ data replicas, a costly and error-prone approach. Contemporary enterprises face a fundamental challenge: enabling organization-wide data sharing across heterogeneous analytics platforms. When $n$ datasets must be accessible from $m$ different query engines, traditional data-warehouse-centric approaches can create up to $n \times m$ replicas, resulting in three critical problems: (1) expansion of data replicas, where multiple departments create redundant copies of the same dataset, increasing storage costs and synchronization overhead; (2) duplicated access configuration, requiring repeated environment-specific setup for each dataset-engine combination; and (3) lack of data consistency, where update timing differences cause discrepancies across query engines accessing logically identical data and schema/permission drift across replicas.

The root cause lies in the limited interoperability between different data warehouses and analytics platforms. For example, BigQuery and Snowflake, two widely adopted cloud data warehouses, cannot directly access each other's native tables. Organizations using both platforms must export data from one system, transform it, and import it into the other, creating duplicate copies and operational overhead. While Python-based machine learning environments offer flexible data access through various client libraries, they generally do not bridge the interoperability gap between SQL-based data warehouses. This fundamental limitation forces organizations into costly and error-prone data replication workflows.

Existing data management approaches offer partial solutions but fail to fully address the multi-query engine challenge. Data lake~\cite{sawadogo2021data} provide cost-effective storage for diverse data formats but require moving data into separate analytics platforms for interactive querying. Data warehouses~\cite{kimball2013data} enable high-performance SQL queries but assume data resides within vendor-specific storage, creating silos when multiple warehouses coexist. Data lakehouses~\cite{armbrust2021lakehouse} combine the flexibility of data lakes with transactional capabilities, yet practical implementation patterns for federated access across heterogeneous query engines remain insufficiently established.

\begin{figure*}[t]
  \centering
  \includegraphics[width=\textwidth]{compare.png}
  \caption{\textbf{From write-many silos to Write-Once, Read-Anywhere.}
  \emph{(a)} In a warehouse-centric setup, each source dataset $s_i$ is copied into each client/engine $c_j$, yielding up to $n\times m$ replicas $d_{ij}$ and siloed reads.
  \emph{(b)} EDSP keeps one canonical table $t_i$ per source and exposes it to all engines via federated access, enabling Write-Once, Read-Anywhere; adding a new engine requires no data migration. Notation $S$, $D$, $T$, and $C$ follows the panel labels.}

  \label{fig:traditional-vs-edsp}
\end{figure*}

To address these limitations, we propose the \textbf{Enterprise Data Science Platform (EDSP)}, a unified data management architecture built on the \textit{Write-Once, Read-Anywhere} principle. EDSP consists of two core design elements:

\begin{itemize}
    \item \textbf{Write-Once}: Each dataset is written once to a single location in object storage as the single source of truth. Updates and appends are performed in place without creating replicas.
    \item \textbf{Read-Anywhere}: Any query engine in the organization can directly access and query the data through federated, in-place access mechanisms, without materializing full copies in engine-specific storage.
\end{itemize}

EDSP implements this principle through a four-layer architecture: (1) Data Preparation Layer for standardized ETL processing, (2) Data Store Layer using data lakehouse on object storage, (3) Access Interface Layer providing unified metadata and documentation, and (4) Query Engine Layer supporting multiple analytics environments. Figure~\ref{fig:traditional-vs-edsp} contrasts traditional data-warehouse-centric approaches with the proposed EDSP architecture, illustrating how EDSP reduces redundant data copies while preserving multi-engine access.

We deployed EDSP in an enterprise production environment and conducted a comprehensive evaluation across three dimensions. Our evaluation suggests that, among the baselines we tested, EDSP was the only solution we observed to achieve federated in-place access across BigQuery, Snowflake, and Python environments. Compared with conventional approaches requiring data migration, EDSP reduces operational steps by 33--44\% for cross-warehouse data sharing. While query latency increases by up to a factor of 2.6 compared with native tables, end-to-end completion times remain within seconds, maintaining practical performance for analytical workloads.

This paper makes the following contributions:

\begin{itemize}
    \item Write-Once, Read-Anywhere principle: A design principle enabling federated, in-place data access from multi-query engine environments to a single data source, eliminating the data replication problem inherent in traditional approaches.
    \item EDSP architecture: A four-layer architecture with clear separation of concerns, enabling addition of new query engines without data migration.
    \item Production validation: Empirical evidence from enterprise deployment demonstrating 33--44\% reduction in operational effort while maintaining practical query performance.
\end{itemize}

The remainder of this paper is organized as follows. Section~\ref{sec:architecture} presents the EDSP architecture and design principles. Section~\ref{sec:implementation} describes the implementation details and technical challenges. Section~\ref{sec:evaluation} evaluates EDSP's effectiveness through three experiments and Section~\ref{sec:discussion} discusses implications and practical considerations. Section~\ref{sec:limitations} addresses limitations and future work. Section~\ref{sec:related} reviews related work, and Section~\ref{sec:conclusion} concludes.

\section{Proposed Architecture}
\label{sec:architecture}
\begin{figure}[t]
  \centering
  \includegraphics[width=\columnwidth]{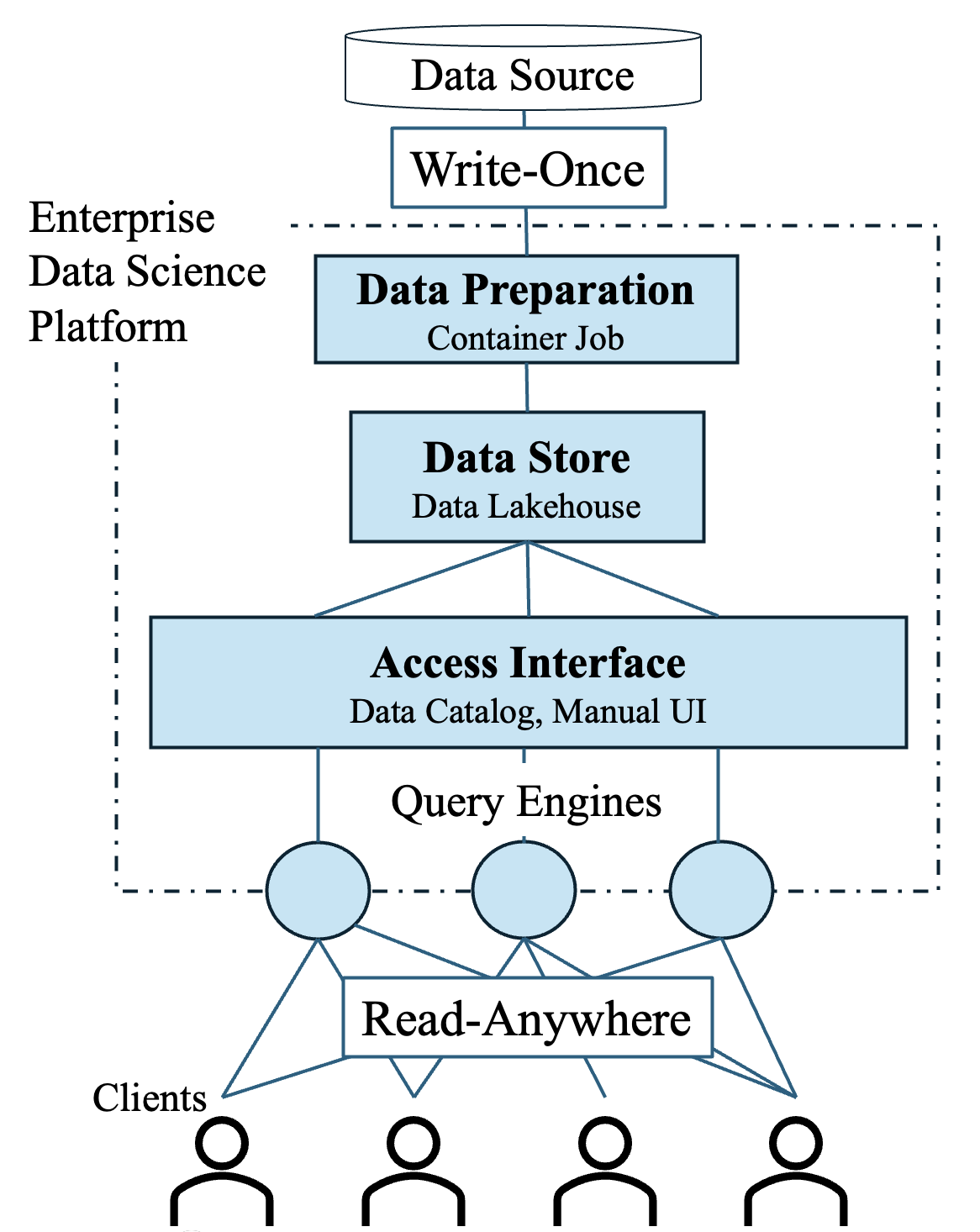}
  \caption{Four-layer architecture of the Enterprise Data Science Platform (EDSP).}
  \label{fig:edsp-architecture}
\end{figure}

This section presents design guidelines for the Enterprise Data Science Platform (EDSP) grounded in the Write-Once, Read-Anywhere principle. As illustrated in Figure~\ref{fig:edsp-architecture}, EDSP adopts a four-layer architecture that enables federated in-place access from diverse analytical environments while eliminating data duplication. We first establish core design principles (Section~\ref{subsec:design_principles}), then detail each architectural layer (Section~\ref{subsec:four_layers}), and describe inter-layer interactions (Section~\ref{subsec:interactions}).

\subsection{Design Principles}
\label{subsec:design_principles}

To apply Write-Once, Read-Anywhere at enterprise scale, we adopt three supporting principles.

Principle 1: Single Source of Truth. 
Each dataset is written once to a designated path in object storage, which serves as the authoritative version; all appends, updates, and schema changes occur in place. Consequently, for $n$ datasets and $m$ query engines, the total number of durable replicas is at most $n$.
In practice, analytical workflows inevitably produce intermediate or derived datasets. EDSP does not forbid such tables. The “write-once” principle is applied at the level of each dataset managed by the platform, meaning that source datasets are written once and then shared across engines.

Principle 2: Federated In-Place Access. Query engines access data directly from object storage without materializing full copies in engine-specific storage. We define \textit{federated in-place access} as the ability for each engine, once configured, to retrieve the current version of a dataset with a single query, reading only the necessary data from object storage and creating no persistent replicas. This is achieved through external table mechanisms (for SQL engines) and direct file I/O via client libraries (for programming environments).

Principle 3: Layer Independence. 
Responsibilities are partitioned across architectural layers and coupled via stable interfaces. Adding a new query engine is a configuration-only change in the Access Interface Layer; data in object storage and other layers remain unchanged. This independence enables evolutionary system design where
technology choices can be updated without full platform
migration.

To put these principles into practice, we adopt the following implementation guidelines:

\begin{enumerate}
    \item Open table formats: Use vendor-neutral formats (Apache Iceberg, Delta Lake) to avoid lock-in to proprietary catalog systems.
    \item Infrastructure as Code: Manage all configurations including access controls, external table definitions, and metadata schemas in version-controlled repositories to ensure reproducibility.
    \item Separation of concerns: Clearly delineate data preparation, storage, access interface, and consumption responsibilities to localize the impact of changes.
\end{enumerate}

\subsection{Four-Layer Architecture}
\label{subsec:four_layers}
As shown in Figure~\ref{fig:edsp-architecture}, EDSP consists of four layers. Each layer has a well-defined responsibility. Data flows unidirectionally from upstream sources through preparation and storage layers to downstream consumption.

\paragraph{Data Preparation Layer}
The first layer ingests and standardizes data from external sources. Implemented as containerized jobs, it transforms raw inputs into analysis-ready tabular formats and writes them to the Data Store layer. This process is executed once per data source, and subsequent updates follow the same pipeline.

\paragraph{Data Store Layer}
The Data Store Layer serves as the organization’s \emph{single source of truth}. Following the data lakehouse pattern, it stores data in object storage while maintaining schema and metadata in a metastore. Open table formats such as Apache Iceberg provide ACID transactions and time travel capabilities, ensuring data consistency across heterogeneous query engines.

\paragraph{Access Interface Layer}
The Access Interface Layer provides a unified interface to the Data Store. It consolidates multiple access modalities by exposing a data catalog for metadata visibility and discovery, generating engine-specific connection details, and documenting access procedures. Users can thus access data using methods best suited to their environments.

\paragraph{Query Engine Layer}
The Query Engine layer comprises the analytical tools used across the organization. SQL-based tools access the Data Store via external table mechanisms, while programming environments use client libraries/ SDKs to read files directly. Adding a new query engine does not require migrating or duplicating existing data.

\subsection{Inter-Layer Interactions}
\label{subsec:interactions}
Data flow proceeds unidirectionally from upstream to downstream layers—specifically, from data sources, through the Data Preparation Layer, into the Data Store. Data placed in the Data Store via the Data Preparation Layer are updated in place, which clarifies data lineage and makes the impact of updates predictable.

Query engines access the Data Store through two standard paths exposed by the Access Interface Layer:
\begin{itemize}
  \item For SQL query engines: access via external table mechanisms.
  \item For programming environments: direct file reads via client libraries/SDKs.
\end{itemize}
Both paths reference the same physical data, thereby realizing Read-Anywhere.

\subsection{Scalability and Vendor Neutrality}
EDSP delivers scalability along two axes. With respect to data volume, scales approximately linearly in our setting due to the characteristics of object storage. With respect to the number of query engines, onboarding new engines typically requires only adding documentation in the Access Interface Layer. New analytical tools can thus be introduced without impacting existing data or engines.

By adopting open table formats, EDSP avoids dependence on specific vendors. For example, if an organization migrates from Snowflake to BigQuery, the Data Store Layer remains unchanged; only configurations in the Access Interface Layer need adjustment. This property preserves long-term flexibility in technology choices.

\section{Implementation}
\label{sec:implementation} 
This section describes the implementation of EDSP, focusing on technology stack selection, layer-specific implementation details, and solutions to key technical challenges encountered during deployment. In this implementation, we used Apache Iceberg on object storage as the core of the Data Store Layer, enabling concurrent access from multiple query engines, including BigQuery, Snowflake, and Python-based workflows.

\subsection{Implementation Overview}
We implemented the four-layer architecture with the following technology stack:
\begin{itemize}
  \item Data Preparation Layer: Container-based job execution platform
  \item Data Store Layer: Apache Iceberg tables on object storage
  \item Access Interface Layer: Web-based documentation platform
  \item Query Engine Layer: Snowflake, BigQuery, Python-based workflows
\end{itemize}

\subsection{Implementation Details of Each Layer}
\paragraph{Data Preparation Layer}
The Data Preparation Layer is implemented as a set of containerized batch jobs. Each job is managed in Git and deployed through a CI/CD pipeline. The jobs ingest data from external sources, standardize schemas, convert data to the Iceberg format, and write both initial loads and incremental updates to the same tables in the same object storage. This realizes the Write-Once aspect of the Write-Once, Read-Anywhere principle.

\paragraph{Data Store Layer}
For the Data Store Layer, we selected Apache Iceberg. Its open-source, vendor-neutral table format mitigates vendor lock-in, and it supports time travel and cross-engine interoperability. A major implementation challenge was ensuring a consistent mechanism for federated data access to the same Iceberg table across multiple data warehouse; in our implementation, BigQuery and Snowflake. Integration generally follows two paths. The first routes through each engine’s proprietary catalog. This can hinder cross‑engine access and may introduces vendor lock-in in the Data Preparation Layer, and was therefore not adopted. The second defines external tables that point to fixed metadata locations. However, Iceberg emits timestamped metadata files whose names change with each commit, so any fixed location becomes stale over time. To resolve this, the Data Store Layer maintains a fixed-name pointer file (\texttt{latest.metadata.json}), independent of catalog-managed metadata, that always references the most recent metadata file. This keeps Iceberg’s object storage metadata as the single source of truth. Importantly, the standard catalog-managed metadata remains available and unmodified; therefore, core Iceberg semantics and catalog-based integrations continue to function as designed.

\paragraph{Access Interface Layer}
The Access Interface Layer is implemented as a web-based documentation platform. For each dataset, we provide metadata such as schema definitions, update frequency, and semantic descriptions, together with engine-specific access methods. Specifically, we maintain version-controlled documents that include external table DDLs for Snowflake, configuration procedures for BigQuery, and code snippets for Python. This layer enables data consumers to easily identify the appropriate access method for their environment.

\paragraph{Query Engine Layer}
We implemented and validated three primary query engines. Snowflake uses the Iceberg external table feature with catalog integration to automatically synchronize metadata. BigQuery defines Iceberg tables via the BigLake external table feature and accesses the latest data by periodically refreshing metadata. In Python-based workflows, we use the PyIceberg library to load data directly into DataFrames with predicate pushdown.

\subsection{Implementation Challenges and Solutions}
During implementation, we encountered the following primary challenges that required careful design decisions

\paragraph{Metadata Compatibility}
Different data warehouse products impose different requirements on metadata files. After evaluating options such as symbolic links and periodic copying, we extended the Iceberg writer to concurrently update a fixed-name pointer file on each metadata change.

\paragraph{Unified Access Control}
Because each query engine has its own authentication mechanism, unifying authentication and authorization was challenging. We addressed this by using object storage Cloud IAM roles as a common basis and mapping them to engine-specific credentials in the Access Interface Layer.

\subsection{Verification of Design Principles}
We verified that the implemented system satisfies the proposed design principles. We evaluated Write-Once and Read-Anywhere independently, as detailed below.

\paragraph{Write-Once}
For seven sample datasets, we performed an initial ingestion followed by periodic updates over the course of one month. We confirmed that all writes targeted the same Iceberg tables in the same object storage, with no auxiliary tables or per-engine copies created.

\paragraph{Read-Anywhere}
We executed equivalent queries against the same datasets from the three primary query engines—Snowflake, BigQuery, and Python-based workflows—and confirmed that the results were consistent across engines. We also demonstrated that onboarding a new query engine required only configuration in the Access Interface Layer to access the existing data, with no data duplication or rewriting.

\paragraph{Federated In-place Access}
By monitoring network traffic and storage usage on the query-engine side, we confirmed that only the data required by query execution was transferred and that no persistent replicas (materialized copies) were created.

\section{Evaluation}
\label{sec:evaluation} 
This section evaluates the effectiveness of EDSP from three perspectives: adherence to the Write-Once, Read-Anywhere principle, reduction in operational effort for data sharing, and query performance characteristics. Our goal is to demonstrate that EDSP resolves the data silo problem in  multi-query engine environments at a production-ready level.

\subsection{Evaluation Setup}
\subsubsection{Experimental Environment and Target Systems}
We conducted the experiments on Google Cloud Platform (GCP) and Amazon Web Services (AWS). BigQuery, Snowflake, Google Cloud Storage (GCS), and Amazon Simple Storage Service (S3) were all placed in the Tokyo region to minimize the impact of network latency.

As baselines, we selected the following three data management approaches currently used in enterprises:
\begin{itemize}
  \item Hugging Face Dataset Hub: a widely adopted sharing platform among machine learning teams.
  \item BigQuery native tables: the standard option in Google Cloud environments.
  \item Snowflake native tables: a major cloud data warehouse with multi-cloud support.
\end{itemize}

BigQuery and Snowflake are representative cloud data warehouses. BigQuery native tables are a standard choice in Google Cloud that enables high-speed analytical queries through columnar storage and distributed processing. Snowflake native tables are a major cloud data warehouse adopted by many enterprises due to their multi-cloud support and automatic scaling. Hugging Face Dataset Hub enables internal data sharing while preserving compatibility with public models, supporting workflows that move from prototyping on public datasets to productionization on proprietary data. It is also a common choice as part of the data platform in organizations with in-house R\&D teams.

\subsubsection{Evaluation Dataset and Queries}
We used the Foursquare POI dataset\footnote{https://opensource.foursquare.com/os-places/}. POI data are a typical cross-functional dataset leveraged for trade-area analysis by Marketing, store search by Product, and location-model training by R\&D. We chose this dataset because it is high-quality and open source with commercial use permitted, and because its size is moderate for our study. In practical operations we filter the dataset to Japan (4{,}785{,}257 rows, 1.35\,GB, 28 columns); therefore, our experiments also target Japanese POIs only.

We selected three query patterns frequently used in practice: Q1 (full extraction, assuming ETL or batch processing), Q2 (retrieval with a WHERE predicate filtering by prefecture with approximately 2.1\% selectivity), and Q3 (GROUP BY aggregation computing category-wise statistics). These patterns represent common analytical workloads across different business functions. Unless otherwise noted, all queries cap output at 10,000 rows.

\subsection{Adherence to Write-Once, Read-Anywhere}
\label{subsec:eval1}
We assess adherence to the Write-Once, Read-Anywhere principle by examining whether data can be directly referenced without creating persistent replicas. Configurations enabling direct access via external tables or official APIs are marked as feasible, while those requiring export/import or format conversion are marked as infeasible.

Table~\ref{tab:write_once_read_anywhere} shows that EDSP is the only solution enabling direct access from all query engine environments. Native tables in BigQuery and Snowflake are mutually inaccessible, which is a primary cause of data silos within organizations. While Python can access each warehouse natively, it does not provide warehouse‑to‑warehouse interoperability. By adopting the Apache Iceberg format, which is officially supported as an external table from both BigQuery and Snowflake and as a direct read API from Python, EDSP achieves Write-Once, Read-Anywhere.

\begin{table}[t]
\centering
\caption{Adherence to the Write-Once, Read-Anywhere principle.}
\label{tab:write_once_read_anywhere}
\begingroup
\setlength{\tabcolsep}{10pt}
\renewcommand{\arraystretch}{1.12}
\begin{tabular}{lcccc}
\toprule
& \multicolumn{4}{c}{\textbf{Data store}} \\
\cmidrule(lr){2-5}
\textbf{Query Engine} & \textbf{HF} & \textbf{BQ} & \textbf{SF} & \textbf{EDSP} \\
\midrule
BigQuery  & -- & $\checkmark$ & -- & $\checkmark$ \\
Snowflake & -- & -- & $\checkmark$ & $\checkmark$ \\
Python    & $\checkmark$ & $\checkmark$ & $\checkmark$ & $\checkmark$ \\
\bottomrule
\end{tabular}
\endgroup

\vspace{0.25em}
\begin{tablenotes}[flushleft]\footnotesize
\item Symbols: $\checkmark$ = direct, federated in-place access; -- = migration or format conversion required.
\item Abbreviations: HF = Hugging Face Dataset Hub, BQ = BigQuery, SF = Snowflake, EDSP = Enterprise Data Science Platform.
\end{tablenotes}
\end{table}

The results reveal that conventional approaches create fragmentation: BigQuery users cannot access Snowflake data without exporting and importing, and vice versa. This forces organizations into costly data replication workflows. EDSP reduces this fragmentation by providing a single data location accessible from all engines, thereby implementing the core Write-Once, Read-Anywhere principle.

\subsection{Reduction in Data-Access Effort}
\label{subsec:eval2}
We evaluate the operational effort required for data access by measuring the number of steps from dataset preparation to first-query execution. This metric captures the workload imposed on data engineers and analysts when using shared data across different analytical environments.

\subsubsection{Methodology: LLM-as-a-Judge}
To objectively assess operational steps across diverse technology stacks, we adopted an LLM-as-a-Judge~\cite{zheng2023judging} methodology. This approach offers three advantages. First, LLMs possess broad knowledge of multiple technology stacks and are less biased toward specific vendors compared to human experts with specialized expertise. Second, it has become common practice in data engineering to utilize LLMs for workflow planning, making this evaluation reflective of real-world practices. Third, using multiple LLMs improves reproducibility and reduces the impact of individual model biases. In addition, we verified that the sets of steps produced by the LLMs are consistent with the official procedures and documentation for each platform.

The 12 step types evaluated include authentication configuration, permission grants, driver or SDK installation, external table registration, query execution, data export, format conversion, transfer via intermediate storage, schema mapping, SQL dialect rewriting, data import, and Iceberg metadata reading. Step necessity was judged by three large language models: GPT-5, Gemini 2.5 Pro, and Claude 4 Sonnet. We assigned equal weight to all steps, a conservative setting for EDSP. EDSP steps are fewer and simpler; assigning equal weights ignores step complexity and likely underestimates EDSP’s advantage.
A single prompt template with two template variables---data source and query engine---was used in all experiments. In each run, the model selected applicable items from a 12-step migration checklist, generated Python scripts and SQL queries to perform the migration, and then verified which checklist steps had been satisfied. Across all models tested, the reported step counts were all within ±1 step of the mean.

\subsubsection{Results}
Table~\ref{tab:step-count} reports the average number of operational steps across the three LLMs. EDSP reduces the number of steps required for data sharing and consumption across multi-query engines.

\begin{table}[t]
\centering
\caption{Operational steps to first query (average across three LLMs).}
\label{tab:step-count}
\begingroup
\setlength{\tabcolsep}{10pt}      
\renewcommand{\arraystretch}{1.15}
\begin{tabular}{l *{4}{S}}
\toprule
& \multicolumn{4}{c}{\textbf{Data store platform}} \\
\cmidrule(lr){2-5}
\textbf{Query Engine} & {\textbf{HF}} & {\textbf{BQ}} & {\textbf{SF}} & {\textbf{EDSP}} \\
\midrule
\textbf{BigQuery}  & 8 & 3 & 9 & 6 \\
\textbf{Snowflake} & 9 & 9 & 3 & 5 \\
\textbf{Python}    & 3 & 4 & 3 & 3 \\
\bottomrule
\end{tabular}
\endgroup

\vspace{0.2em}
\small\emph{Abbreviations: HF = Hugging Face Dataset Hub; BQ = BigQuery; SF = Snowflake; EDSP = Enterprise Data Science Platform.}
\end{table}

When the query engine is Snowflake and the data source is BigQuery, the step count is 9 with conventional approaches but 5 with EDSP, a 44\% reduction. Similarly, when the query engine is BigQuery and the data source is Snowflake, the step count decreases from 9 to 6 with EDSP, a 33\% reduction. 

For source-engine pairs that require data migration in conventional approaches, operators must perform a sequence of data movement steps including export, format conversion, transfer via intermediate storage, and import. In EDSP, these are replaced by external table registration and Iceberg metadata reading, enabling the observed reduction. Moreover, the tasks required by EDSP at each step are simpler than those of the other approaches. For example, external table registration in EDSP is a single DDL statement, whereas data export and import require coordinating multiple tools and monitoring long-running jobs.

The reduction in operational steps scales with organizational growth. The eliminated steps primarily involve data movement, which is time-consuming, error-prone, and difficult to automate. By replacing these with metadata operations, EDSP simultaneously simplifies workflows and improves reliability.

\subsection{Query Performance}
\label{subsec:eval3}
We measured query performance to assess the practical overhead of federated data access compared to native table storage. Understanding this trade-off is critical for evaluating EDSP’s suitability for production workloads.

\subsubsection{Measurement Method}
\label{subsub:evalset}
We measured query latency under cold-start conditions without caching, representing the worst-case scenario where no prior queries have warmed up the engine's internal caches. For each query pattern (Q1, Q2, Q3), we executed 30 independent runs and report the median (p50) and 95th percentile (p95) latencies. Python timings include full client-side download and are not directly comparable to server-side SQL latencies. All measurements were conducted with BigQuery, Snowflake, and Python environments accessing either native tables (as baselines) or EDSP's Iceberg tables. Snowflake was configured with an X-Small virtual warehouse, and experiments on GCP Compute Engine used e2-standard-4 instances.

Table~\ref{tab:latency} compares access from each query engine to its native data sources versus access from each query engine to EDSP. This design quantifies the performance overhead of introducing EDSP. Combinations such as Snowflake accessing BigQuery native tables are technically infeasible and thus excluded, as already discussed in Section~\ref{subsec:eval1}. For the Python-based environment, we include Hugging Face Dataset Hub as an additional baseline to verify practicality in machine learning workflows. In this evaluation, we therefore focus on two access patterns, copying data into each engine and querying it as native tables, and querying EDSP’s Iceberg tables directly from each engine without replication. Other multi-query engine configurations in our environment internally rely on such replication before running native queries and are treated as variants of the replication-based baseline.

\begin{table*}[t]
\centering
\caption{Query execution latency in seconds. We report median (p50) and 95th percentile (p95) under cold-start conditions.}
\label{tab:latency}
\begingroup
\setlength{\tabcolsep}{8pt}      
\renewcommand{\arraystretch}{1.18}
\begin{tabular}{l *{6}{S}}
\toprule
 & \multicolumn{2}{c}{\textbf{Q1: Full extraction}}
 & \multicolumn{2}{c}{\textbf{Q2: WHERE predicate}}
 & \multicolumn{2}{c}{\textbf{Q3: GROUP BY}} \\
\cmidrule(lr){2-3}\cmidrule(lr){4-5}\cmidrule(lr){6-7}
\textbf{Configuration} & {\textbf{p50}} & {\textbf{p95}} & {\textbf{p50}} & {\textbf{p95}} & {\textbf{p50}} & {\textbf{p95}} \\
\midrule
Snowflake SQL $\rightarrow$ Snowflake Native Tables    & 1.04 & 1.60 & 0.73 & 0.85 & 0.42 & 0.48 \\
Snowflake SQL $\rightarrow$ Enterprise Data Science Platform      & 2.20 & 3.20 & 1.65 & 2.15 & 1.15 & 1.35 \\
\midrule
BigQuery SQL $\rightarrow$ BigQuery Native Tables     & 3.26 & 3.82 & 3.28 & 3.52 & 1.20 & 1.35 \\
BigQuery SQL $\rightarrow$ Enterprise Data Science Platform       & 4.89 & 6.28 & 6.03 & 6.80 & 3.17 & 4.20 \\
\midrule
Python $\rightarrow$ Hugging Face Dataset Hub & 9.09 & 9.53 & 9.23 & 9.75 & 9.12 & 9.86 \\
Python $\rightarrow$ Enterprise Data Science Platform & 7.69 & 9.76 & 8.95 & 9.42 & 9.32 & 9.87 \\
\bottomrule
\end{tabular}
\endgroup

\vspace{0.25em}
\small\emph{Cold-start runs with caches disabled; 30 repetitions per query; same dataset and region as in Section~\ref{subsub:evalset}.}
\end{table*}

\subsubsection{Results}

The results are shown in Table~\ref{tab:latency}. Accessing data via EDSP increases latency by up to a factor of 2.6 (p50; BigQuery on Q3) compared with native tables, with the maximum overhead observed for BigQuery on the Q3 query. This overhead stems from three sources: reading Iceberg metadata to determine which data files to access, fetching data from object storage through external table functionality, and the lack of engine-specific optimizations such as BigQuery's native columnar format or Snowflake's micro-partitions.

Nevertheless, all queries complete within 10 seconds at p95, which remains within a practical range for analytical workloads. Data science tasks typically involve exploratory queries with human-in-the-loop analysis, where response times on the order of seconds are acceptable.

Considering the benefits of reduced data duplication (33--44\% fewer operational steps, as shown in Section~\ref{subsec:eval2}) and vendor neutrality (federated in-place access from multiple engines, as shown in Section~\ref{subsec:eval1}), this latency increase represents an acceptable trade-off for organizations prioritizing flexibility and operational efficiency over raw query speed. For frequently accessed datasets requiring minimal latency, organizations can employ hybrid strategies such as materializing hot data in native tables while maintaining the canonical version in EDSP for less time-sensitive access patterns.

\section{Discussion}
\label{sec:discussion}

Our results show that EDSP removes copy-heavy workflows while preserving practical performance for its target workloads. By exposing one canonical table per dataset to multi-query engines via federated in-place access, EDSP achieves Read-Anywhere interoperability (Section~\ref{subsec:eval1}) and reduces the steps needed to make data available across engines (Section~\ref{subsec:eval2}). The trade-off is additional latency relative to native storage; in our measurements the worst median increase is up to a factor of 2.6, yet p95 latencies remain in the seconds range (Section~\ref{subsec:eval3}). For exploratory analytics and periodic batch jobs, this balance favors EDSP because it simplifies operations without breaking interactive use.

EDSP is well suited to organizations with multi-query engine analytics that expect periodic engine introductions or replacements. Integrating a new engine required only updates to the Access Interface Layer, specifically configuration and documentation, and did not necessitate data migration assuming compatibility with Apache Iceberg. While detailed limitations and future work are discussed later, our evaluation supports EDSP as a practical choice for multi-query engine analytics when flexibility and reduced operational effort outweigh raw query speed.

\section{Limitations and Future Work}
\label{sec:limitations}

While EDSP demonstrates practical effectiveness in addressing data silos across multi-query engine environments, this study has several limitations that warrant discussion. We organize these limitations into two categories: evaluation scope, experimental constraints.

\subsection{Evaluation Scope}
Our evaluation focused on the data-access effort reduction and cross-engine access, but did not assess following enterprise adoption criteria.

Long-term maintainability remains unverified; issues such as Iceberg table compaction frequency, metadata-driven performance drift, and backward-compatible schema evolution were out of scope. A one-month production deployment suggests operational viability but cannot characterize long-term behavior.

Our total cost of ownership analysis is also limited. Production costs, including object storage API charges, multi-cloud egress, and compute for external tables, are not fully quantified. Early modeling suggests EDSP's advantage grows with more query engines but may erode with frequent reads under per-scan or metered compute pricing. This implies a context-dependent competitiveness threshold not yet validated; EDSP can be cost-efficient with high engine diversity but may exceed conventional approaches in some cases. A cross-organizational cost-benefit analysis remains future work, and adopters should run pilots to measure workload- and pricing-specific costs.

Finally, we did not evaluate organizational factors. Clear data ownership, cross-departmental cost allocation for shared infrastructure, and change management during migration from existing warehouses are critical but out of scope. Prior work on data governance~\cite{khatri2010designing} underscores that technology must be paired with processes and accountability aligned to the architecture.

\subsection{Experimental Constraints}

The experiments were conducted under idealized conditions and do not fully reflect the complexity of real-world environments. The dataset we used (1.35 GB) is sufficient for many analytical tasks, but systems may exhibit different behavior at larger scales. In particular, Apache Iceberg, the storage layer in EDSP, scales to TBs with tuned layouts and regular maintenance~\cite{jain2023cidr}. Performance under concurrent, heterogeneous queries across multiple engines remains unclear: planning and pruning, metadata contention, and maintenance cadence may differ, so cost and latency at TB scale are uncertain. We leave a systematic evaluation for future work.

Because the evaluation was limited to a single region (Tokyo), we did not assess increased latency in multi-region deployments or compliance with data residency requirements. Many global enterprises must replicate data across geographic regions to meet regulatory constraints or to reduce latency for distributed user populations. EDSP's architecture supports multi-region deployment through object storage replication, but the operational complexity and consistency guarantees of cross-region Iceberg tables require further investigation. Additionally, network egress costs in multi-region configurations can be substantial, potentially affecting total cost of ownership calculations.

Finally, our evaluation of the LLM-as-a-Judge~\cite{zheng2023judging} methodology for measuring operational steps represents a novel approach that has not been extensively scrutinized in the literature. Although inter-LLM agreement was high and human expert validation showed strong concordance, the methodology's sensitivity to prompt phrasing and potential biases across different technology stacks warrant further investigation. We view this as an opportunity for future methodological research rather than a limitation of EDSP itself, but acknowledge that alternative evaluation methods (e.g., time-and-motion studies with practitioners) would provide complementary evidence.

\subsection{Future Work}

Future work should address the study’s evaluation limits by establishing long-term maintainability under sustained production conditions beyond the one-month pilot. A complete total-cost analysis is also required: total compute, storage and network cost should be modeled and validated across organization sizes and cloud providers, yielding competitiveness thresholds as engine diversity and read frequency vary. In parallel, organizational prerequisites such as clear data ownership warrant systematic study, informed by data governance practice.

Experimental limitations also motivate broader validation. TB-scale, concurrent, multi-query engine environments should be evaluated to establish practical cost–latency viability. Multi-region deployments should be assessed to measure latency and the consistency of cross-region tables, confirm data-residency compliance, and factor egress into total cost of ownership. The LLM-as-a-Judge method should be evaluated for robustness to prompt variation and model-specific biases. Its findings should be corroborated by practitioner studies, such as time-and-motion measures of task completion time.

Despite these limitations, this study presents a practical, deployable solution with empirical evidence of effectiveness for a clearly defined problem: reducing the effort required to access data in environments with multiple query engines and avoiding data duplication by establishing a single source of truth. The design principles, architectural patterns, and implementation techniques presented here provide actionable guidance for organizations pursuing data democratization across heterogeneous analytics platforms.

\section{Related Work}
\label{sec:related} 
We review related work in three areas: data management architectures, enterprise data management approaches, and open table formats. For each area, we clarify how EDSP differs from existing approaches and addresses gaps in prior work.

\subsection{Data Management Architectures}
Three primary architectural patterns have emerged for enterprise data management: data lakes, data warehouses, and data lakehouses. Each offers distinct trade-offs between flexibility,
performance, and interoperability.
\paragraph{Data Lake \cite{sawadogo2021data}}
Data lakes enable cost-effective storage for data in diverse formats. They employ a schema-on-read approach that retains raw data and interprets structure at query time. However, challenges remain in multi-query engine environments. For example, direct access from BigQuery to an AWS S3–based data lake is not natively supported, necessitating cross-cloud data movement. Additionally, insufficient metadata management impedes data discovery and quality assurance.

\paragraph{Data Warehouse \cite{kimball2013data}}
Data warehouses provide fast query performance and consistency by fixing the schema at ingestion time (schema-on-write). Nevertheless, sharing data across heterogeneous data warehouse systems is difficult. To share data between BigQuery and Snowflake, a full data migration is typically required, and bi-directional references via external tables are not well established. As a result, when an organization operates multiple data warehouses, data duplication and silos often arise.

\paragraph{Data Lakehouse \cite{armbrust2021lakehouse}}
Data lakehouses integrate the flexibility of data lakes with the transactional capabilities of data warehouses. They bring ACID properties to object storage and enable fast SQL queries. However, behavior in multi-query engine environments can be inconsistent; for instance, read access by data warehouses may depend on the specific catalog implementation. Moreover, implementation patterns for federated access are still maturing for building an enterprise-scale data hub on lakehouse technologies, and empirical evaluation in production environments remains limited.

\subsection{Approaches to Enterprise Data Management}
Several conceptual frameworks have been proposed to address enterprise-wide data management challenges, each emphasizing different aspects of data sharing and governance.

\paragraph{Enterprise Data Marketplace\cite{Eichler2023}}
An Enterprise Data Marketplace (EDMP) is a platform aimed at democratizing data within an enterprise. Beyond cataloging, it integrates access control and provisioning to support the end-to-end workflow from data discovery to consumption. However, EDMP primarily focus on metadata and access control and do not prescribe concrete mechanisms for enabling data access across heterogeneous query engines. In addition, consumer environments are often constrained to conform to platform-imposed specifications.

\paragraph{Data Mesh \cite{machado2022data}}
Data Mesh is a decentralized approach that emphasizes domain autonomy. Each domain manages data as a product and shares it via a self-service platform, while computable governance seeks to preserve consistency in a distributed setting. However, it provides limited guidance on concrete technical implementation and does not clearly address interoperability across heterogeneous query engines.

\paragraph{Data Fabric \cite{blohm2024data}}
Data Fabric is a concept that automates data integration by leveraging active metadata and knowledge graphs. It uses AI to optimize data discovery, integration, and delivery. That said, the definition is abstract and implementation approaches remain vague. Practical application to multi-query engine environments is largely not addressed.

\subsection{Open Table Formats}

Open table formats provide data warehouse–equivalent capabilities on object storage. Representative implementations include Apache Iceberg \cite{icebergSpec}, Delta Lake \cite{armbrust2020delta}, and Apache Hudi \cite{hudiTechSpec}.

Among these, Iceberg currently offers the broadest ecosystem support. Major data warehouses such as BigQuery and Snowflake can read Iceberg tables via external table features, as can query engines including Spark. Each format differs in how it maintains metadata and in its engine support; the most appropriate choice depends on the workload \cite{jain2023cidr}. While OTFs show promise for next-generation data management, remaining vendor dependencies are a concern, and empirical adoption and evaluation remain limited.

Engines such as Presto and Trino, and catalog services such as Databricks Unity Catalog, provide SQL access and governance for data stored in systems including BigQuery and Snowflake. Presto and Trino can query BigQuery and Snowflake via connectors like JDBC, but they do not enable data access originating from BigQuery or Snowflake to other data warehouses. In our environment, Unity Catalog offers only limited options for using BigQuery as a federated access point to external systems. By contrast, EDSP exposes a write-once, read-anywhere data store in open table formats that can be accessed directly from multiple warehouse engines, including BigQuery and Snowflake.

\section{Conclusions}
\label{sec:conclusion} 
This paper proposed the Enterprise Data Science Platform (EDSP), a unified data management architecture grounded in the Write-Once, Read-Anywhere principle, to address data management challenges in multi-query engine environments. The heterogeneity of departmental analytical tools, for instance BigQuery for business intelligence, Snowflake for data warehousing, and Python for machine learning, leads to data duplication and operational overhead, which are critical challenges. When $n$ datasets must be accessible from $m$ query engines, traditional approaches can create up to $n \times m$ replicas, leading to inconsistencies, increased costs, and management complexity.

Through implementation and evaluation in a production environment, we demonstrated the following results. First, based on the baselines we evaluated, EDSP was the only solution we found that supports federated in-place access across BigQuery, Snowflake, and Python environments, validating the Write-Once, Read-Anywhere principle. Second, for data sharing across heterogeneous query engines and data sources, EDSP reduced operational steps by 33–44\% compared with conventional approaches requiring data migration. The eliminated steps primarily involved time-consuming, error-prone data movement; EDSP replaces them with straightforward, low-overhead external table definitions and metadata-only operations. Third, although a latency increase of up to a factor of 2.6 was observed relative to native tables, all queries completed within 10 seconds at the 95th percentile, maintaining practical performance for analytical use cases.

The contributions of this work extend beyond demonstrating technical feasibility. By shifting data sharing from bulk data movement to predominantly metadata‑ and authorization‑based access, EDSP lowers interdepartmental barriers to data utilization and promotes data-driven decision-making across organizations. The adoption of Apache Iceberg as an open format preserves flexibility for future technology choices, enabling organizations to add new query engines or migrate between vendors without data replication. The design principles and implementation patterns presented here provide actionable guidance for organizations pursuing data democratization in heterogeneous analytics environments.

Future work includes performance validation on TB-scale datasets, comprehensive total cost of ownership analysis across diverse organizational scales, integration with real-time streaming workloads, and evaluation in multi-region deployments. Despite these remaining challenges, EDSP demonstrates that the Write-Once, Read-Anywhere principle can be realized in production environments, offering a practical solution to the long-standing problem of data silos in multi-query engine enterprises.

\newpage
\bibliographystyle{IEEEtran}
\bibliography{ref}

\end{document}